\documentstyle[aps,prl,amssymb,epsfig]{revtex}
\begin{document}
\draft
\twocolumn[\hsize\textwidth\columnwidth\hsize\csname %
@twocolumnfalse\endcsname

\title{Spin-dependent thermoelectric transport coefficients in near-perfect
quantum wires}
\author{T. Rejec$^1$ and A. Ram\v sak$^{1,2}$}
\address{$^1$J. Stefan Institute, 1000 Ljubljana, Slovenia}
\address{$^{2}$ Faculty of Mathematics and Physics, University of
Ljubljana, 1000 Ljubljana, Slovenia }
\author{J.H. Jefferson}
\address{QinetiQ, Sensors and Electronic Division,
St. Andrews Road, Great Malvern,\\
Worcestershire WR14 3PS, England}
\date{\today}
\maketitle
\begin{abstract}\widetext
Thermoelectric transport coefficients are determined for semiconductor
quantum wires with weak thickness fluctuations.
Such systems exhibit anomalies in conductance near 1/4 and 3/4 of
$2e^2/h$ on the rising edge to the first conductance plateau,
explained by singlet and triplet resonances of conducting electrons
with a single weakly bound electron in the wire (T. Rejec,
A. Ram\v sak, and J.H. Jefferson, Phys. Rev. B {\bf 62}, 12985 (2000)).
We extend this work to study the Seebeck thermopower coefficient and
linear thermal conductance within the framework of the
Landauer-B\" uttiker formalism, which also exhibit anomalous structures.
These
features are generic and robust, surviving to temperatures of a few
degrees. It is shown quantitatively how at elevated temperatures
thermal conductance progressively deviates from the Wiedemann-Franz
law.
\end{abstract}
]
%%%%%%%%%%%%%%%%%%%%%%%%%%%%%%%%%%%%%%%%%%%%%%%%%%%%%%%%%%%%%%%%%%%%%%%%
\narrowtext
\section{Introduction}

One of the main properties of small confined electron systems,
intensively studied experimentally and theoretically in the last decade,
is the electrical conductance. However, other transport coefficients
also serve as a sensitive probe of new phenomena in such systems, such
as the thermopower of chaotic quantum dots \cite{godijn99} or of
atomic size metallic contacts \cite{ludoph99} and most recently,
anomalies in one-dimensional wires \cite{appleyard00}. Theoretical
investigations predict in these systems a range of new properties of
transport coefficients, such as anomalously enhanced thermopower in
quantum dots due to the Kondo effect \cite{kim01} and, at low
temperatures, changes in sign together with linear thermal
conductance violating Wiedemann-Franz law \cite{boese01}. Anomalies in
thermoelectric coefficients are also found in standard strongly
correlated systems: the Anderson model \cite{hewsonbook}, the Hubbard
model \cite{stafford93} and the $t$-$J$ model \cite{jaklic}.

In this paper, we extend our recent theoretical study of conductance
anomalies to include thermoelectric effects due to a temperature
gradient. Anomalies are related to weakly bound electron states
within the quantum wire. In particular, we consider a small
fluctuation in thickness of the wire in some region giving rise to a
weak bulge. If this bulge is very weak then only a single electron
will be bound. We may thus regard this system as an `open' quantum dot
in which the bound electron inhibits the transport of conduction
electrons. Near the conduction threshold, there is a 'Coulomb
blockade' and we have shown that this gives rise to
spin-dependent resonances, also in an axial magnetic field, for wires of
both
rectangular \cite{rejec002d} and cylindrical \cite{rejec003d}
cross-section.

Experimentally, the staircase structure of the conductance of quantum
wires was reported more than a decade ago \cite{wees88}, and more
recent systematic investigations showed unexpected structure in the
rising edge to the first conductance plateau
\cite{thomas96,kane,kristensen98,kaufman99}.

Here we model a quantum wire as in
Ref.~\cite{rejec003d} and, explicitly, we assume a wire of circular
symmetry about the $z$-axis with constant potential, $V(r,z)=0$ within
a boundary $r_{0}(z)$ from the symmetry axis and confining potential
$V_{0}>0$ elsewhere. This geometry is close to that of narrow
'v'-groove quantum wires, which also exhibit anomalies near the
conductance threshold\cite{kaufman99}. To be definite, we choose
parameters appropriate to GaAs for the wire and Al$_{x}$Ga$_{1-x}$ As
for the barrier with $x$ such that $V_{0}=0.4$eV, which is close to
the crossover to indirect gap. The wire width is taken as
$r_{0}(z)=\frac{1}{2}a_{0}( 1+\xi \cos ^{2}\pi z/a_{1}) $ for $|z|\leq
\frac{1}{2}a_{1}$ and $r_{0}(z)\equiv \frac{1}{2}a_{0}$ otherwise,
i.e., a wire of width $a_{0}$ with a single bulge of length $a_{1}$
and width $(1+\xi )a_{0}$, as shown in insets of Fig.~1(c) and
Fig.~2(c).

\section{Conductance}

We consider the interacting electron problem with the above wire
thickness variation in a range which ensures that only one electron
occupies a bound state and that restriction to a single channel near
the conduction edge is an excellent approximation. This is always the
case for a very weak smooth variation, i.e. a near perfect wire.  From
numerically exact solutions of the two-electron scattering problem,
the conductance is calculated from our generalisation of the usual
Landauer-B\"{u}ttiker (LB) formula \cite{landauer57}, to include
spin-dependent scattering \cite{oppenheimer} of conduction electrons
from the single electron bound in the potential well. This gives
$G(\mu)=G_{0}\;{\cal T}(\mu)$, where $G_{0}=2e^{2}/h$, $\mu$ is the
Fermi energy and the transmitivity is a weighted average over singlet
and triplet channels \cite{rejec002d,rejec003d,flambaum00},
\begin{equation}
{\cal T}(\mu)=\frac{1}{4}{\cal T}_{\mathrm s}(\mu)
+\frac{3}{4}{\cal T}_{\mathrm t}(\mu)\label{tau}.
\end{equation}
At elevated temperatures we use the LB finite temperature extension
\begin{equation}
G(\mu)=G_0\int\biggl[-\frac{\partial f(\epsilon,\mu,T)}
{\partial \epsilon }\biggr] {\cal T}(\epsilon )d\epsilon,
\label{T}
\end{equation}
where $f(\epsilon,\mu,T)=(1+\exp [(\epsilon-\mu) /k_{ {\rm
B}}T])^{-1}$ is the usual Fermi function which describes the thermal
distribution of electrons in the leads.  $G(\mu)$ is shown in
Fig.~1(a) and Fig.~2(a) for a wire with relatively small and a larger
bulge, respectively. Here the energy is measured from the threshold of
the conductance. As discussed in Ref.~\cite{rejec003d}, the weak
bulge in the wire is equivalent to a shallow potential well in a
perfectly straight wire and if the length of the bulge region is
small, this effective potential well can only accommodate one bound
state with the consequence that only a singlet resonance in $G$
exists, as observed, for example, in Ref.~\cite{kaufman99}.
Conversely, if the bulge region is longer, both, singlet and triplet
resonances contribute.  For even longer bulge regions with a very
shallow effective potential well (near perfect wire), the singlet
resonance is pushed to lower energy and therefore becomes extremely
narrow. In this regime, only the broader triplet can be resolved at
finite temperature \cite{rejec002d,rejec003d}, as observed
experimentally in clean gated structures
\cite{thomas96,kane,kristensen98}.

\section{Thermoelectric effects}

The LB approach can be extended to include electrical and heat
currents through a region between two leads with different
temperatures and chemical potentials \cite{sivan86,proetto91}.  With
$T+\Delta T$, $\mu+eU$ for the left lead and $T$, $\mu$ for the right
lead, we get
\begin{eqnarray}
j&=&\frac{2e}{h}\int\Delta f(\epsilon) {\cal T}(\epsilon){\mathrm
d}\epsilon,\\
j_{\mathrm Q}&=&\frac{2}{h}\int(\epsilon-\mu)\Delta f(\epsilon)
{\cal T}(\epsilon){\mathrm d}\epsilon,
\end{eqnarray}
and
\begin{equation}
\Delta f(\epsilon)=f(\epsilon,\mu+eU,T+\Delta
T)-f(\epsilon,\mu,T).\label{deltaf}
\end{equation}
In the linear response regime of vanishing $\Delta T$ and $U$ the
currents simplify to
\begin{eqnarray}
j&=&\frac{2e^2}{h}K_0(\mu)U+\frac{2e}{h}K_1(\mu)\frac{\Delta
T}{T},\label{j}\\
j_{\mathrm Q}&=&\frac{2e}{h}K_1(\mu)U+\frac{2}{h}K_2(\mu)\frac{\Delta T}{T},
\end{eqnarray}
where
\begin{equation}
K_n(\mu)=-\int(\epsilon-\mu)^n
\frac{\partial f(\epsilon,\mu,T)}
{\partial \epsilon }
{\cal T}(\epsilon){\mathrm d}\epsilon.
\label{k}
\end{equation}

\begin{figure}[hbt]
\center{\epsfig{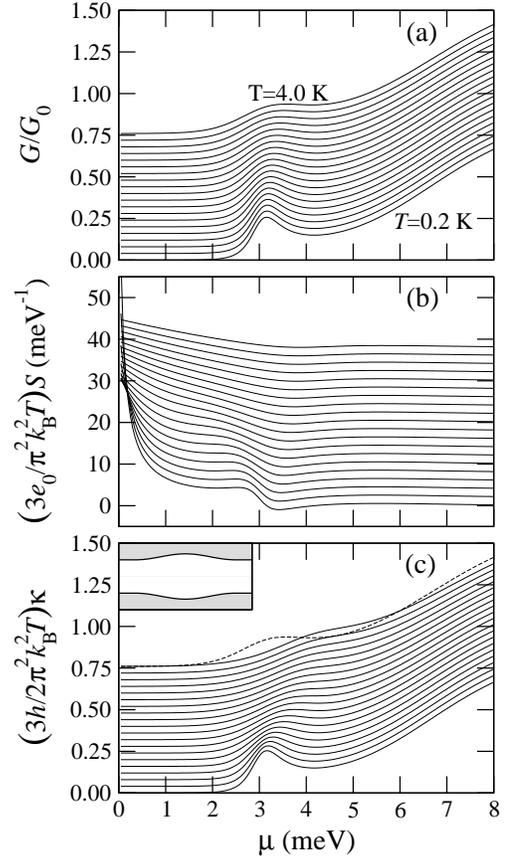}}
\caption{(a) Electrical conductance $G(\mu)$, (b) thermopower $S(\mu)$,
and (c) thermal
conductance $\kappa(\mu)$ for wire parameters $a_0=10$~nm,
$a_1=30$~nm, $\xi=0.18$ and screening length
$\rho=100$~nm. Other parameters and the
numerical method is as in Ref.~10. %cite{rejec003d}
The dashed line in (c) represents Wiedemann-Franz law result for
$T=4$K. The traces for different $T$ are offset vertically for clarity.}
\end{figure}

\subsection{Thermopower}

The Seebeck thermopower coefficient $S$ measures the voltage
difference needed to neutralize the current due to the temperature
difference between the leads. In the linear response regime the
thermopower is given by,
\begin{equation}
S(\mu)=\frac{U}{\Delta T}=-\frac{1}{eT}\frac{K_1(\mu)}{K_0(\mu)},
\label{cm}
\end{equation}
as is for various systems discussed in
Refs.~\cite{proetto91,appleyard00}. Eq.~\ref{cm} is formally the same
as the Mott-Jones formula for simple metals \cite{mott36} and generalized
for a system with stronger electron-phonon interactions in
Refs.~\cite{jonson}.

\begin{figure}[hbt]
\center{\epsfig{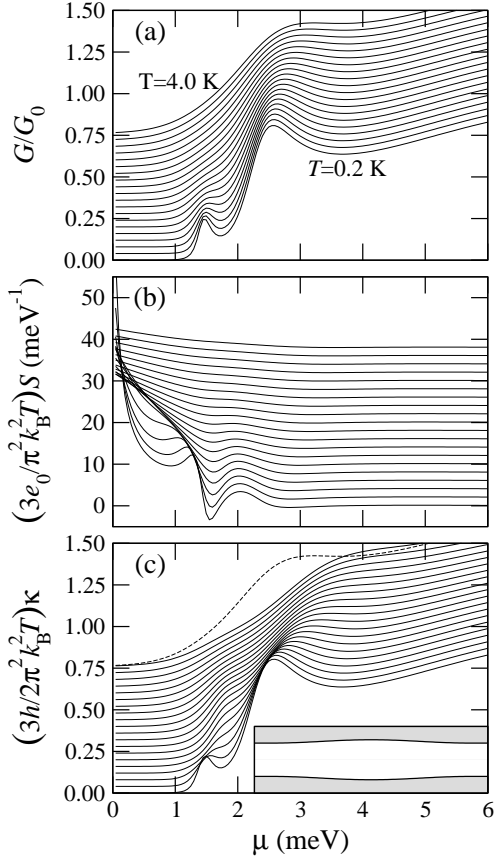}}
\caption{As Fig.1, but for longer bulge with parameters
$a_0=10$~nm, $a_1=60$~nm and $\xi=0.1$.}
\end{figure}

In Fig.~1(b) the thermopower of a narrow wire with a small bulge is
presented for the same range of temperatures as $G(\mu)$.  Such a
result is expected, e.g., for the system studied in
Ref.~\cite{kaufman99}.  The structure reflects the singlet resonance
observed in the conductance and is smeared out at temperatures
comparable with the width of the resonance. In a wire with small
thickness variation, but with a longer bulge, triplet resonance
scattering also exists, as shown in Fig.~2. In the thermopower curve
of Fig.~2(b), the dominant structure at lower temperatures comes from
the singlet resonance, though the triplet resonance is still clearly
discernible. At higher temperatures the triplet structure is washed
out first, in contrast to the conductance result, Fig.~2(a).  At low
temperatures only the transmitivity at energies close to the chemical
potential contributes to the above integrals and the general result
Eq.~\ref{cm} can be related to the temperature dependent $G(\mu)$ by
the following expansion
\begin{equation}
S(\mu)=-\frac{\pi^2{k_{\mathrm B}}^2T}{3e}\biggl(\frac{\partial
\ln G(\mu)}{\partial
\mu}+
\frac{\pi^2 k_{\mathrm B}^2T^2}{15 \,G(\mu)} \frac{\partial^3
G(\mu)}{\partial
\mu^3}\biggr)\!+ ... \label{slin}
\end{equation}
Our results were calculated using the exact relation
Eq.~\ref{cm}. However, the leading term in Eq.~\ref{slin}, is a
reasonable approximation for energies above the singlet resonance and
up to temperatures where the structure is thermally smeared out.  This
is shown in Fig.~3(a) where we present a comparison of $S(\mu)$ for the
exact result with the approximations to first and second order.  We
see that at energies below the resonance, both the linear and cubic
approximations deviate significantly from the exact result,
Eq.~\ref{cm}. In this regime the conductance is itself very small and
hence $G(\mu)^{-1}\partial^nG(\mu) / \partial \mu^n $ is prone to
error making calculations and experimental data analysis based on
this expansion unreliable.

The thermopower of one-dimensional wires has been measured
\cite{appleyard98,molenkamp90} and more recently, further anomalies
related to `0.7 anomaly' in conductance were reported
\cite{appleyard00}. The authors of Ref.~\cite{appleyard00} observe a
dip in $S(\mu)$ at energies corresponding to the anomaly in $G(\mu)$.
However, the logarithmic derivative with respect to the gate voltage
of the measured $G$ exhibits a much deeper minimum than the dip in the
measured $S(\mu)$, which remains well above zero even at the lowest
temperatures . This clearly shows that a simple non-interacting
formula is not valid in this low temperature regime.
Apart from the small corrections to the logarithmic approximation to
$S$, our model and its solution within the LB framework are
\begin{figure}[hbt]
\center{\epsfig{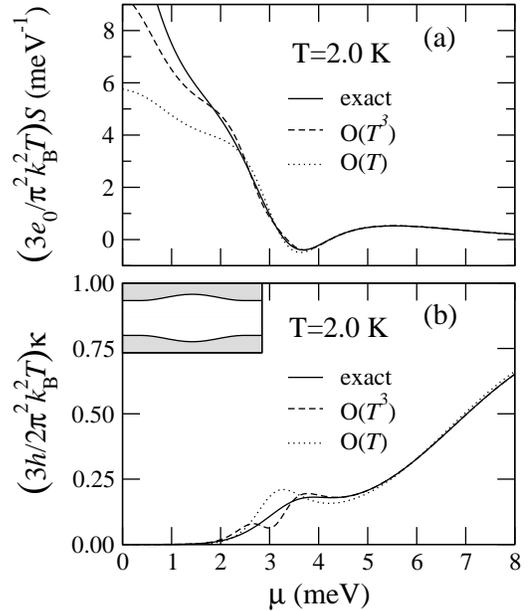}}
\vskip 0.4 cm
\caption{(a) Thermopower as obtained with Eq.~\ref{cm}
for $T=2$K and parameters used for Fig.~1 (full line). Dashed and dotted
lines
correspond respectively to the result of
Eq.~\ref{slin} and the linear $T$ approximation (first term in
Eq.~\ref{slin}).
(b) Thermal conductance -- parameters as in (a).}
\end{figure}

\noindent
in agreement with the findings of Ref.~\cite{appleyard00}. That is,
the calculated thermopower is in good agreement with experiment except
at low temperatures where we also predict a deep minimum. This
discrepancy at low-temperatures may well be a many-body Kondo-like
effect contained within our model but not within the two-electron
approximation we have used here and in our earlier papers. We
expect the two-electron approximation to break down at low
temperatures for which the underlying extended Hubbard model, which is
the starting point of our approach, can be mapped onto a generalised
Anderson model with coupling terms that are strongly energy dependent
\cite{rejecand}. The standard results for the single impurity problem
\cite{hewsonbook} cannot be applied directly to this effective model,
which is the subject of current research \cite{rejecram}. At very low
temperatures, a Kondo-like resonance is expected \cite{boese01}, for
which many-body effects would dominate with a breakdown of
formula Eq.~\ref{cm}.

\subsection{Thermal conductance}
The linear thermal conductance is the heat current divided by the
temperature difference between the leads when the chemical potentials
are adjusted to give no electrical current. From Eqs.~\ref{j}-\ref{k}
we see that this is related to ${\cal T}(\epsilon)$ by,
\begin{equation}
\kappa(\mu)=\frac{2}{h T}\left(K_2(\mu)-\frac{K_1^2(\mu)}{K_0(\mu)}\right).
\label{kapa}
\end{equation}
For low temperatures this simplifies to Wiedemann-Franz
law, first term in
\begin{eqnarray}
\label{wf}
\kappa(\mu)&=&\frac{\pi^2k_{\mathrm B}^2T}
{3e^2}G(\mu)\biggl(1 + \\
&+&\frac{\pi^2k_{\mathrm B}^2T^2}{15}\biggl[
\frac{8}{G(\mu)}\frac{\partial^2G(\mu)}{\partial\mu^2}
-5\bigl(\frac{\partial \ln G(\mu)}{\partial \mu}\bigr)^2
\biggr]\biggr)
+... \nonumber
\end{eqnarray}

In Fig.~1(c) and Fig.~2(c) $\kappa(\mu)$ is shown for $T$ from 0.2K
to 4K, calculated from Eq.~\ref{kapa}.  Comparison of Figs.~1(a),2(a)
with Figs.~1(c),2(c) shows good agreement with the Wiedemann-Franz law
at lower temperatures but there is increasing deviation at higher
temperatures in the resonance region. For comparison, the dashed lines
in Fig.~1(c), Fig.~2(c) show the corresponding linear approximation result,
Eq.~\ref{wf}. This is also seen in the plot of $\kappa$ for $T=2$K
shown in Fig.~3(b). One of the most striking features of these plots is that
$\kappa(\mu)$, calculated from Eq.~\ref{kapa}, exhibits an anomaly at
higher energies than the corresponding anomaly in conductance, a
prediction which is open to experimental verification.

\section{Summary}

In summary we have, within the framework of the LB approach,
calculated thermal transport coefficients for near-perfect quantum
semiconductor quantum wires, extending our earlier work on
spin-dependent conduction anomalies.
These anomalies are a
universal effect in one-dimensional systems with very weak
longitudinal confinement. The emergence of a specific structure
$G(\mu)\sim \frac{1}{4}G_{0}$ and $G\sim\frac{3}{4}G_{0}$ is a spin
effect, being a direct consequence of the singlet and triplet
nature of the resonances. The probability ratio 1:3
for singlet and triplet scattering follows directly from this and as
such is a universal effect, not only for conductance but all thermoelectric
transport coefficients. A comprehensive numerical investigation of open
quantum dots using a wide range of parameters shows that singlet resonances
are always at lower energies than the triplets, in accordance with the
Lieb-Mattis theorem for bound states \cite{lieb62}.

Thermopower plots show anomalies, related ultimately to the Coulomb
interaction between a localised electron and the remaining conduction
electrons. We have shown that the lower-energy singlet anomalies in
thermopower are more pronounced. These should be clearly observable in
wires which show the corresponding conductance anomalies, such as the
narrow `hard confined' wires reported in Ref.~\cite{kaufman99}, or in
gated quantum wires under high source-drain bias where the singlet
anomaly is clearly observed \cite{patel91}.

Finally we conclude by emphasising that although our model of a
quantum wire with a weak bulge may appear rather specialised, it is
actually quite general since the weak bulge is mathematically
equivalent to a weak potential well in an otherwise perfect wire. As
with our previous work, we have not investigated in detail the actual
causes of such weak effective (or real) potential wells but point out
that they may well be due to quite different sources in different
experiments, e.g. thickness fluctuations, remote impurities or gates,
electronic polarisation, or some other more subtle electron
interaction effect. The main point is that because the effective
potential well is shallow, {\em it will bind one and only one
electron}. The universal anomalies in conductance and thermopower are
a direct consequence of this and occur for a wide range of
circumstances in almost perfect quantum wires.

\section{Acknowledgments}
The authors wish to acknowledge N. Appleyard and M. Pepper for drawing our
attention to the thermopower problem in quantum wires and for helpful
discussions. This work was supported by MESS, the EU and the UK MoD.

\end{document}